\documentclass[letterpaper, 10 pt, conference]{ieeeconf}  

\IEEEoverridecommandlockouts                              

\usepackage{multicol} 
\usepackage{multirow}
\usepackage{color} 							
\usepackage{pgfplots} 					
\usepackage{tikz} 							
\usepackage{amsmath,bm,amsfonts} 
\usepackage{graphicx,epstopdf}	
\usepackage{mathtools} 
\usepackage{pdfpages}						
\usepackage{hyperref} 					
\usepackage{subfigure}

\usepackage{float}
\usepackage{array}
\usepackage{nicefrac}
\newcolumntype{L}[1]{>{\raggedright\let\newline\\\arraybackslash\hspace{0pt}}m{#1}}
\usepackage{centernot}

\usepackage{acronym}

\usepackage{verbatim}

\usepackage{cuted}

\usepackage{cite}

\usepgfplotslibrary{fillbetween}
\usetikzlibrary{patterns}

\usetikzlibrary{spy}

\usetikzlibrary{pgfplots.groupplots}

\newtheorem{define}{Definition}
\newtheorem{assumption}{Assumption}

\makeatletter
\newcommand\Autoref[1]{\@first@ref#1,@}
\def\@throw@dot#1.#2@{#1}
\def\@set@refname#1{
	\edef\@tmp{\getrefbykeydefault{#1}{anchor}{}}%
	\xdef\@tmp{\expandafter\@throw@dot\@tmp.@}%
	\ltx@IfUndefined{\@tmp autorefnameplural}%
	{\def\@refname{\@nameuse{\@tmp autorefname}s}}%
	{\def\@refname{\@nameuse{\@tmp autorefnameplural}}}%
}
\def\@first@ref#1,#2{%
	\ifx#2@\autoref{#1}\let\@nextref\@gobble
	\else%
	\@set@refname{#1}
	\@refname~\ref{#1}
	\let\@nextref\@next@ref
	\fi%
	\@nextref#2%
}
\def\@next@ref#1,#2{%
	\ifx#2@ and~\ref{#1}\let\@nextref\@gobble
	\else, \ref{#1}
	\fi%
	\@nextref#2%
}

\hypersetup{nolinks=true}

\newcommand{\bb}[1]{\mathbb{#1}} 
\newcommand{\bmat}[1]{\begin{bmatrix} #1 \end{bmatrix}} 
\newcommand{\cone}{\mbox{cone}} 

\newcommand{\Lcal}{\mathcal{L}}

\acrodef{iid}[i.i.d.]{independent, identically distributed}

\newcommand{\defeq}{\vcentcolon=}

\title{\LARGE \bf
Issues with Input-Space Representation in Nonlinear Data-Based Dissipativity Estimation
}

\author{Ethan J. LoCicero$^{1}$ and Alexander Penne$^{1}$ and Leila Bridgeman$^{1}$
\thanks{This work is supported by NSF GRFP Grant No. 1644868, the Alfred P. Sloan Foundation, ONR Grant No. N00014-23-1-2043, and NSF Grant No. 2303158. Corresponding author: Ethan J. LoCicero.}%
\thanks{$^{1}$Ethan J. LoCicero (Research Associate) and Leila Bridgeman (Assistant Professor) are with the Dept. of Mechancial Eng. and Materials Science at Duke Univeristy, Durham, NC, 27708, USA (email {\tt\small ejl48@duke.edu}; {\tt\small ljb48@duke.edu}, phone 919-660-5310). Alexander Penne (Undergraduate) is with the Dept. of Electrical and Computer Engineering at Duke Univeristy (email {\tt\small alexander.penne@duke.edu}).}%
}

\begin{document}

\maketitle
\thispagestyle{empty}
\pagestyle{empty}

\begin{abstract}
In data-based control, dissipativity can be a powerful tool for attaining stability guarantees for nonlinear systems if that dissipativity can be inferred from data. This work provides a tutorial on several existing methods for data-based dissipativity estimation of nonlinear systems. The interplay between the underlying assumptions of these methods and their sample complexity is investigated. It is shown that methods based on $\delta$-covering result in an intractable trade-off between sample complexity and robustness. A new method is proposed to quantify the robustness of machine learning-based dissipativity estimation. It is shown that this method achieves a more tractable trade-off between robustness and sample complexity. Several numerical case studies demonstrate the results.
\end{abstract}

\section{Introduction} \label{sec:Introduction}
There has been significant interest recently in data-based control, where a control policy is either designed directly from input-output data, or a plant model is identified from data to inform controller design \cite{Zhao2024}. Many of these methods rely on linearity of the unknown system to attain stability guarantees with relatively little data. For nonlinear systems, attaining such guarantees is more challenging \cite{Martin2023}. 

In model-based control, identifying a dissipative characterization of the nonlinear system is one approach to assuring stability. Dissipativity is a input-output property that generalizes gain, passivity, and conic sectors, among others. Given a dissipative plant, the Dissipativity Theorem \cite{Vidyasagar1981} provides tractable constraints on the open-loop properties of a controller that guarantee closed-loop stability. This constraint can used to recover robust stability in optimal control problems based on a nominal linearization \cite{Geromel1997a}. There are many model-based tools for characterizing dissipativity \cite{Hill1980}. Characterizing dissipativity directly from data would enable robust design strategies for data-based control.

If the dissipativity characterization identified from a finite data set is guaranteed to hold for all possible system trajectories, the it is said to be ``robust". This robustness is critical for guaranteeing stability through the Dissipativity Theorem. Robust dissipativity estimation methods for linear systems based on Willems' Fundamental Lemma \cite{Willems2005} are well-developed and have data requirements as low as a single trajectory from a persistently exciting input \cite{Koch2022}. For nonlinear systems, achieving robustness is more difficult and requires much more data. Some approaches have relaxed the problem to allow for a partially known model or full state access \cite{Martin2021,Martin2023TAC}. This work reviews methods for off-line robust dissipativity estimation of unknown nonlinear systems with only input-output data \cite{Montenbruck2016,Romer2017,RomerGaussian,Tang2019,Tang2019journal,Tang2021}.

The contribution of this work is primarily tutorial. Section~\ref{sec:l2} clarifies the assumptions that underlie existing methods for robust nonlinear dissipativity estimation and investigates the consequences of those assumptions on the claimed robustness. Section~\ref{sec:delta} then reviews a class of ``$\delta$-covering" methods \cite{Montenbruck2016,Romer2017,RomerGaussian} and provides a modest extension thereof. It is demonstrated that for these methods, either the sample complexity is too high to be implemented, or the desired robustness property does not hold as expected. Section~\ref{sec:ML} reviews a class of methods based on machine learning techniques \cite{Tang2019,Tang2019journal,Tang2021}. A new approach to verify robustness for these methods is proposed based on the generalization error from probably approximately correct learning \cite{ShalevShwartz2014}. It is shown that this generalization error decouples the sample complexity from the assumptions investigated in Section~\ref{sec:l2}, so robustness can be practically achieved given an appropriate data generation method. Several such methods are explored in numerical examples.

\section{Preliminaries} \label{sec:Preliminaries}
The set of strictly positive real numbers is denoted $\bb{R}_>$, and $\bb{R}_\geq \defeq \bb{R}_>\cup \{0\}$. The set of $n$-dimensional real vectors is denoted $\bb{R}^n$. The real and imaginary components of a complex number, $x$, are $\mathrm{Re}(x)$ and $\mathrm{Im}(x)$. The identity matrix is $I$, and $(\cdot)^T$ and $(\cdot)^{-1}$ denote the transpose and inverse of real matrices. If $x\in\bb{R}$, then $|x|$ denotes the absolute value. If $\mathcal{S}$ is a set, then $|\mathcal{S}|$ is its cardonality. A function $f:\bb{R}^n\rightarrow\bb{R}$ has complexity $\mathcal{O}(g(x))$, denoted $f(x)\sim\mathcal{O}(g(x))$, if there exists $k,x_0\in\bb{R}_>$ and $g:\bb{R}^n\rightarrow\bb{R}$ such that $|f(x)|\leq kg(x)$ for all $x\geq x_0$. Conversely, $f(x)\sim\Omega(g(x))$ if $|f(x)| \geq kg(x)$ for all $x\geq x_0$. This provides a lower bound on complexity. Time, space, and sample complexity refer to the amount of computations, storage, and samples required to execute an algorithm.

Let $\mathcal{X}$ be a real inner product space with inner product $\langle(\cdot),(\cdot)\rangle:\mathcal{X}\times\mathcal{X}\rightarrow\bb{R}$ and induced norm $\|x\|\defeq\sqrt{\langle x,x\rangle}<\infty$. If the elements of $\mathcal{X}$ are $n$-dimensional vector sequences, i.e. $x:\bb{R}\rightarrow\bb{R}^n$ for all $x\in\mathcal{X}$, then the space is denoted $\mathcal{X}^n$, when relevant. The extension of $\mathcal{X}$, denoted $\mathcal{X}_e$, satisfies $\|x\|_T^2 \defeq \|x_{T}\|^2<\infty$ for all $T\in \bb{R}_\geq$, where $x_{T}$ is the truncation of $x(k)$ at $k=T$, defined as $x(k) = x(k)$ for $k\leq T$ and $x(k) = 0$ for $k>T$. The truncated inner product is $\langle x,y\rangle_T \defeq \langle x_T,y_T\rangle$ for all $x,y\in\mathcal{X}$. The space of square integrable functions is $\mathcal{L}_2$, which has inner product $\int_0^\infty x^Ty dt$. The Frobenius norm and $\mathcal{L}_\infty$ norm are denoted $\|(\cdot)\|_F$ and $\|\cdot\|_{\mathcal{L}_\infty}$, respectively. When not specified, the induced norm of the relevant Hilbert space is assumed. Let $\widehat{u}$ denote the Fourier transform of $u:\mathbb{R}\rightarrow\mathbb{R}^n$. If $\mathcal{G}$ is an LTI system, then $\widehat{\mathcal{G}}(\omega)$ denotes its transfer function.

\begin{define}\label{def:diss}
    (\hspace{-.01mm}\cite{Hill1980}) Let $\mathcal{U}$ and $\mathcal{Y}$ be real Hilbert spaces. An operator $\mathcal{G}:\mathcal{U}_e\rightarrow \mathcal{Y}_e$ is $(Q,S,R)$-dissipative if
    \begin{align}
        \langle y,Qy\rangle_T + \langle y,Su \rangle_T + \langle u,Ru\rangle_T \geq 0 \label{eqn:diss_orig}
    \end{align}
    for all $u\in\mathcal{U}_e$ and $T\in\bb{R}_>$, where $Q$, $S$, and $R$ are real matrices of appropriate dimensions. An operator is \textit{ultimately virtual} $(Q,S,R)$-dissipative if 
    \begin{align}
        \langle y,Qy\rangle + \langle y,Su \rangle + \langle u,Ru\rangle \geq 0 \label{eqn:diss}
    \end{align}
    for all $u\in\mathbf{K}(\mathcal{G}) \defeq \{u\in\mathcal{U} \, | \, y\in\mathcal{Y}\}$. If, in addition, $\mathbf{K}(\mathcal{G}) = \mathcal{U}$, then $\mathcal{G}$ is \textit{ultimately} $(Q,S,R)$-dissipative.
\end{define}
Conic sectors are a special case of dissipativity that are used here for illustrative purposes. An interior conic sector is $\mathrm{cone}_c(r)\defeq (-I,(r^2{-}c^2I,2cI)$-dissipative or $\mathrm{cone}(a,b)\defeq $ $(-I,\frac{a+b}{2}I,-abI)$-dissipative, where $c$, $r$, $a$, and $b$ are the center, radius, lower bound, and upper bound, respectively. Degenerate conic bounds are $\mathrm{cone}(a,\infty) \defeq (0,\frac{1}{2}I,-aI)$ and $\mathrm{cone}(-\infty,b) \defeq (0,-\frac{1}{2}I,b)$ \cite{Zames1966}.
\begin{define}
    A $\delta$-ball around the point $u_i\in\mathcal{U}$ is $\mathcal{B}^\alpha_\delta(u_i) = \{u \; | \; \|u-u_i\|_\alpha \leq \delta\}$, where $\delta\in\bb{R}_>$ is the covering radius, and $\alpha$ indicates the norm ($\Lcal_2$, $\mathcal{L}_\infty$, etc).
\end{define}
\begin{define}
	A collection of points $\{u_i\}_{i=1}^K$ is an $\delta$-covering in $\alpha$-norm for the set $\mathcal{U}$ if $\mathcal{U}\subseteq \cup_{i=1}^K \mathcal{B}^\alpha_\delta(u_i)$.
\end{define}

\section{Representing {$\mathcal{L}_{2e}$}} \label{sec:l2}

From Definition~\ref{def:diss}, an operator, $\mathcal{G}$, is dissipative if and only if Equation~\ref{eqn:diss_orig} holds for all $u\in\mathcal{U}_e$, where $\mathcal{U}_e$ is a Hilbert space defining the set of permissible inputs. For Definition~\ref{def:diss} to have practical utility for robust control, the space $\mathcal{U}_e$ must be rich enough to represent all possible inputs that the system will encounter during operation. This is usually taken to be the space $\mathcal{L}_{2e}$, which is the space of all signals with finite energy over a finite time domain. Verifying \autoref{eqn:diss_orig} for each signal in $\mathcal{L}_{2e}$ independently would require infinite data. This problem is easily circumvented for LTI systems because their behavior can be fully characterized by their response to either a persistently exciting input (via Willems' Fundamental Theorem \cite{Willems2005}) or the set $\mathcal{U}_e = \{\sin(\omega t) \; \forall \; \omega\in\bb{R}_\geq\}$ (as in the classical approach to experimental Nyquist analysis \cite{Nyquist1932}). These simplifications do not hold for nonlinear systems. Therefore, existing methods \cite{Montenbruck2016,Romer2017,RomerGaussian,Tang2021} make several assumptions on $\mathcal{U}_e$. The first two have been explicitly stated in various ways in the literature.
\begin{assumption}
    The amplitude of the permissible inputs is absolutely bounded above by some constant, $\bar{u}\in\bb{R}_>$, i.e. $\|u(t)\|_{\mathcal{L}_\infty}\leq \bar{u}$ for all $t\in\bb{R}_\geq$, $u\in\mathcal{U}_e$.
\end{assumption}
\begin{assumption}
    The induced norm of the permissible inputs is bounded below by some constant, $\epsilon\in\bb{R}_>$, i.e. $\|u\| \geq \epsilon$ for all $u\in\mathcal{U}_e$.
\end{assumption}
The first assumption is justified by the physical limitations of the system actuators and the environment, which cannot generate instantaneously infinite signals. The second is necessary because arbitrarily small input signals cannot be densely sampled. It is also practical for maintaining a sufficient signal-to-noise ratio in data collection. The third assumption below has not been explicitly stated in the literature.
\begin{assumption}
    If \autoref{eqn:diss_orig} holds for one sufficiently large $T\in\bb{R}_>$, then it holds for all $T\in\bb{R}_>$.
\end{assumption}
By setting $T$ much larger than the time scales of interest, Assumption 3 approximates ultimate virtual dissipativity. If $\mathcal{G}$ is causal and $Q$ is negative definite (which encompasses many, but not all, important cases), then ultimate virtual dissipativity implies dissipativity \cite[Theorem 1]{Hill1980}. Therefore, Assumption 3 is often justified as an approximation for dissipativity. Applying these three assumptions to $\mathcal{L}_{2e}$ results in an input space that can be represented with an infinite set of orthonormal basis functions, such as Legendre polynomials, Fourier bases, or wavelets. To make the problem tractable, one last assumption is usually made.
\begin{assumption}
    The set of permissible inputs may be represented by a finite number, $b$, of orthonormal basis functions, $v_1$,$\dots$, $v_b$, i.e. $\forall$ $u\in\mathcal{U}_e$, $\exists$ $\alpha_i\in\bb{R}$ such that $u = \sum_{i=1}^{b} \alpha_i v_i$.
\end{assumption}

This fourth assumption is motivated by the fact that physical systems have a diminishing response to high-frequency signals. Therefore, if the neglected basis functions encode high-frequency information, their impact on the operator's dissipativity is expected to be negligible. Define $\mathcal{U}_{A1234} \defeq \{\mathcal{L}_{2e} \, | \, \mbox{Assumptions 1, 2, 3, and 4}\}$. As $\bar{u}$, $T$, $b\rightarrow\infty$ and $\epsilon\rightarrow 0$, $\mathcal{U}_{A1234}\rightarrow \mathcal{L}_{2e}$. Consequently, the dissipativity properties of a system on $\mathcal{U}_{A1234}$ tend to the system's properties on $\mathcal{L}_{2e}$. Nonetheless, to verify dissipativity on $\mathcal{U}_{A1234}$ there remain infinitely many $u\in\mathcal{U}_{A1234}$ to test.  

The following sections review the major existing strategies to guarantee dissipativity on $\mathcal{U}_{A1234}$ using a finite sample set. Section~\ref{sec:delta} shows that the sample complexity of $\delta$-covering methods is so large that a system's behavior on $\mathcal{U}_{A1234}$ is unlikely to represent its behavior on $\mathcal{L}_{2e}$ with a practical sample size. Then, Section~\ref{sec:ML} shows that machine learning methods can derive probabilistic guarantees of dissipativity on $\mathcal{U}_{A1234}$ without this sample complexity problem.


\section{$\delta$-covering Methods} \label{sec:delta}

\subsection{Summary and Complexity}

In \cite{Montenbruck2016}, an $\mathcal{L}_2$-norm $\delta$-covering of $\mathcal{U}_{A1234}$ is proposed as a way of constructing guaranteed dissipativity properties in the special cases of gain and passivity indices, which was also used for general SISO dissipativity in \cite{Romer2017}. This method requires an additional assumption on the unknown system.
\begin{assumption}
    The operator, $\mathcal{G}$, is Lipschiz continuous, i.e. for some $L>0$, $\|\mathcal{G}(u_1) - \mathcal{G}(u_2)\|\leq L\|u_1-u_2\|$  for all $u_1,u_2,\in\mathcal{U}_e$. Moreover, $L$, or a bound on $L$, is known.
\end{assumption}
With this assumption, \cite{Montenbruck2016} proposes the set of sampled trajectories to be
\begin{align}
    \mathcal{U}' {=} \left\{ u' {=} \sum_{i=1}^{b} \alpha_i v_i \, \Big| \, \left\{ \bar{u}(k-N)/N\right\}_{k=1}^{2N-1} \right\}, \label{eqn:sampling}
\end{align}
where $2N{-}1$ is the number of samples along each basis dimension. This sampling procedure ensures $\mathcal{U}'$ is a $\delta$-cover of $\mathcal{U}_{A1234}$ with a covering radius of $\delta = b\bar{u}/(2N{-}1)$ using $K = (2N{-}1)^{b}$ samples. The resulting dissipativity characterization is conservative, and the exact characterization is achieved at $\delta \rightarrow 0$. Defining the sampling density as $\rho = \bar{u}/\delta$, the number of samples is
\begin{align}
    K = (\rho b)^{b},
\end{align}
so $K\sim\mathcal{O}(b^b)$ for a constant sampling density. Further, $K \sim \Omega(\rho^b)$ for any sampling procedure because it takes $\rho^b$ samples to $\delta$-cover $\mathcal{U}_{A1234}$ in the $\mathcal{L}_\infty$ norm, and $\mathcal{B}_\delta^{\mathcal{L}_2}\subseteq \mathcal{B}_\delta^{\mathcal{L}_\infty}$. Therefore, while more efficient methods than \autoref{eqn:sampling} could be derived for generating a $\delta$-cover, all will have super-exponential sample complexity with respect to the number of bases. This holds even for the probabilistic covering in \cite{Korda2020}. Several different methods of calculating dissipativity have been proposed using this sampling procedure. In \cite{Romer2017}, extreme ray enumeration is used to calculate $(qI,sI,rI)$-dissipativity, but this requires $\mathcal{O}(K)$ space complexity and $\mathcal{O}(K^2)$ time complexity \cite{Avis1992}. In \cite{Montenbruck2016}, special cases of dissipativity including gain, passivity indices, and $\mathrm{cone}_c(r)$ are calculated directly with $\mathcal{O}(1)$ space and $\mathcal{O}(K)$ time complexity. Appendix A extends this method to calculate $\mathrm{cone}(a,b)$ for $-\infty\leq a < b \leq \infty$ with the same complexity.

\subsection{Non-robustness Demonstration}

One consequence of the extreme sample complexity is that $\delta$-covering methods are restricted to systems low-dimensional inputs, as noted in \cite{Tang2021}. What has been less noted is that even for single-input systems, $b$ must be small because $K\sim \mathcal{O}(b^b)$. This is noticeable in the literature, where all existing case studies consider $b\leq 5$ \cite{Montenbruck2016,Romer2017,RomerGaussian,Tang2021}. This section demonstrates that for small $b$, the dissipativity properties guaranteed on $\mathcal{U}_{A1234}$ can be vastly different than the dissipativity properties that hold on $\mathcal{L}_{2e}$.

Specifically, let $\mathcal{G}_L:\mathcal{L}_{2e}\rightarrow\mathcal{L}_{2e}$ be an LTI system with $\|\widehat{\mathcal{G}}_L(\omega_i)\| = \gamma_{\omega_i}$, consider an orthonormal Fourier basis for $\mathcal{U}_{A1234}$ with $b,T,e,\bar{u}<\infty$. For all $u\in\mathcal{U}_{A1234}$, there exists $\alpha_i$ such that $u(t)=\sum_{i=1}^b \alpha_iv_i$, where $v_i = v_i'/\|v_i'\|$, $v_i' = \cos(\omega_it)$, and $\omega_i = \frac{2\pi i}{T}$. Let $\gamma_{max} \defeq \max_i \gamma_{\omega_i}$, $\omega_{max} \defeq \arg\min \gamma_{\omega_i}$ and likewise for $\gamma_{min}$ and $\omega_{min}$. Applying Parseval's Theorem, Triangle Inequality, and Cauchy-Schwartz Inequality, it may be shown that $\|\mathcal{G}u\| = \frac{1}{2\pi}\|\widehat{\mathcal{G}}\widehat{u}\| \leq \sum_{i=1}^b|\nicefrac{\alpha_i}{\|v_i\|}|\gamma_{\omega_i}$ for all $u\in\mathcal{U}_{A1234}$, which is a weighted average of the gain responses at each basis frequency. Therefore, the extremum gain responses for the data set are achieved by sampling each basis independently. Further, if $\omega_{max}$ and $\omega_{min}$ are not sampled basis frequencies, then $\gamma_{max}$ and $\gamma_{min}$ will never be realized in the data. Further, any other orthonormal basis can be represented as a Fourier series, as in $v_i' = \sum_{k=0}^\infty \beta_{i,k}\cos(\omega_k t + \theta_k)$ for some $\beta_{i,k}\geq 0$ and $\theta_k\in\mathbb{R}$. It can then be shown that $\|\mathcal{G}u\| \leq  \sum_{i=1}^b|\frac{\alpha_i}{\|v_i\|}|\left(\sum_{k=0}^\infty |\beta_{i,k}|\gamma_{\omega_k}\right)$. This is a weighted average taken over the frequency components of the basis functions' Fourier series. Again, the extremum values in the data are obtained by sampling each basis independently, and $\gamma_{max}$ is only attained if for some $i$, $\beta_{i,k} = 1$ for $\omega_k=\omega_{max}$ and $0$ otherwise. In all other cases, even if a basis contains a nonzero component of $\omega_{max}$, it is averaged with $\omega_i$ corresponding to smaller gains, so $\gamma_{max}$ is not attained in the data. The same holds for $\gamma_{min}$.

\begin{figure*}
\centering
\input{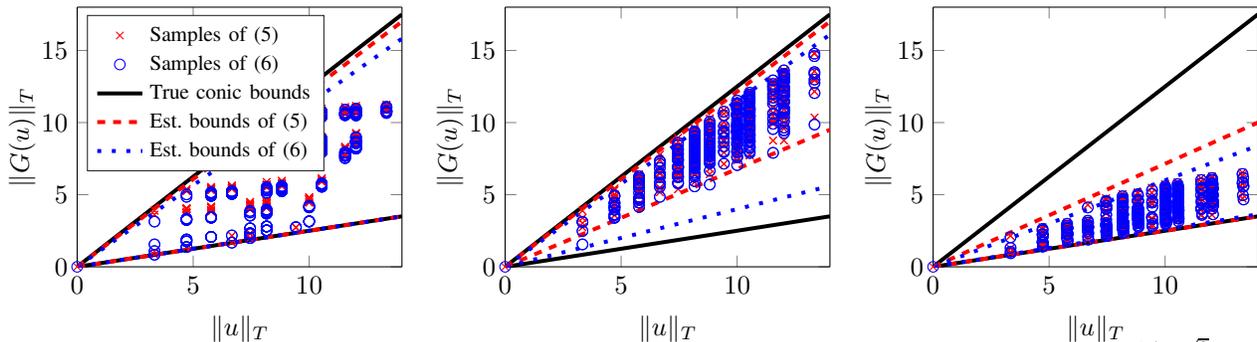}
\vspace{-.5cm}
	\caption{Left: $K=625$ trajectories of Equations~\ref{eqn:example} (LTI) and~\ref{eqn:nonlinear} (nonlinear) generated with a $b=4$ non-sequential Fourier bases $\big\{\frac{1}{T},\,\frac{\sqrt{2}}{T}\sin(\frac{2\pi}{T} t),$ $\frac{\sqrt{2}}{T}\sin(\frac{20\pi}{T}t),\,\frac{\sqrt{2}}{T}\sin(\frac{200\pi}{T}t)\big\}$ with $T=10$. Middle: the same data generated from the first $b=4$ Legendre Polynomial bases with $T=10$, and Right: with $T=1$. Each plot also shows the true conic bounds and those calculated via Appendix A with $\delta=0$, which are nearly identical up to $K=8.1e5$.
 }\label{fig:grid}
\end{figure*}

For illustration, consider the LTI system
\begin{align}
   \widehat{\mathcal{G}}(\omega) = \frac{1}{j\omega+1}+\frac{1}{4}. \label{eqn:example}
\end{align} 
The Nyquist plot of this stable system is a circle with center at $(\nicefrac{3}{4},0)$ and radius $\nicefrac{1}{2}$. Therefore, its tightest dissipative characterization on $\mathcal{L}_{2e}$ is cone$(\nicefrac{1}{4},\nicefrac{5}{4})$. Since the conservatism of $\delta$-covering methods vanishes at $\delta\rightarrow 0$, then for fixed $b$, $\bar{u}$, and $\epsilon$, it is expected that as $K\rightarrow \infty$, the estimated dissipativity of \autoref{eqn:example} on $\mathcal{U}_{A1234}$ will converge to cone$(\nicefrac{1}{4},\nicefrac{5}{4})$. This is not always the case; in fact, it is rarely the case unless the basis functions are chosen very carefully, or if $b$ is very large. For instance, take Equation~\ref{eqn:example} with $b=4$ Fourier bases: $v_1 = \frac{1}{T}$, $v_2 = \frac{\sqrt{2}}{T}\sin(\frac{2\pi}{T} t)$, $v_3 = \frac{\sqrt{2}}{T}\sin(\frac{20\pi}{T}t)$, and $v_4 = \frac{\sqrt{2}}{T}\sin(\frac{200\pi}{T}t)$. The first basis corresponds to $\omega=0$, which is where the Nyquist plot of \autoref{eqn:example} attains the upper conic bound. The last basis approximates $\omega\rightarrow \infty$, which is where the Nyquist plot attains the lower conic bound. The data generated in $\mathcal{U}'$ with these bases are displayed in the rop row of Figure~\ref{fig:grid}, which shows that they cover the entire conic sector well. Correspondingly, the estimated dissipativity will eventually converge to the correct values. This is only possible because the system is known and linear. In general, these choices of basis functions are not necessarily informative. Without exploiting \textit{a priori} information, a common approach is to use the first $b$ Legendre polynomials as basis functions \cite{Montenbruck2016,RomerGaussian}. Figure~\ref{fig:grid} shows the results from this approach with $b=4$ and $T=10$ or $T=1$ in the middle and bottom rows, respectively. The longer time horizon only generates low-frequency data, so it does not appropriately characterize the lower conic bound, instead converging to $\mathrm{cone}(0.68,1.25)$. The shorter time horizon only generates high-frequency data, so it does not appropriately characterize the upper conic bound, instead converging to $\mathrm{cone}(0.25,0.71)$. The short horizon also threatens to violate Assumption 3. Even at $\delta=0$, $K\rightarrow \infty$, neither of these data generation methods can converge to the true dissipativity values. To ensure dissipativity is appropriately characterized for an LTI system using the $\delta$-covering method, a sufficiently large $T$ and $b$ are required. An increase in $T$ reduces the frequency content of the basis functions, so an even larger increase in $b$ is required to compensate, and a linear increase in $b$ results in a super-exponential increase in the number of samples, $K$ to achieve the same level of refinement, $\delta$. 

The same experiment is applied to a pendulum,
\begin{align} \label{eqn:nonlinear}
    \Sigma : \left\{\ddot{\theta} = -\sin(\theta) - \dot{\theta} + u, \qquad y = \dot{\theta} + \nicefrac{1}{4}u\right\},
\end{align}
where the output ensures $\Sigma{\in}\cone(\nicefrac{1}{4},\nicefrac{5}{4})$ by applying the Hamilton Jacobi Inequality with $V {=} (1{-}\cos(\theta)){+}\frac{1}{2}\dot{\theta}^2$ \cite{Schaft1992}. The results in Figure~\ref{fig:grid} show the pattern holds for nonlinear systems too.

A variation on the $\delta$-covering method uses a Gaussian process to represent uncertainty in the dissipativity (specifically, passivity index), and applies Bayesian optimization to sequentially sample regions of $\mathcal{U}_{A1234}$ with the largest uncertainty \cite{RomerGaussian}. This has the potential to significantly reduce the number of samples necessary to achieve the same accuracy. However, it is noted in \cite{RomerGaussian} that this method will similarly struggle from the curse of dimensionality with large $b$, and the computational cost per sample is larger than the $\delta$-covering methods. More work is needed to determine if the Gaussian process method can be extended to more general dissipativity and if it results in sub-exponential sample complexity relative to $b$.

\section{Machine Learning Methods} \label{sec:ML}
An alternative to the $\delta$-covering is to represent dissipativity estimation as a machine learning problem \cite{Tang2019journal,Tang2021}. 
Instead of bounding the worst-case trajectories with a $\delta$-cover or Gaussian process, machine learning methods aim to achieve statistical guarantees that the the data is sufficiently descriptive. A variety of machine learning methods have been proposed \cite{Tang2021}; however, in most cases, error bounds and complexity analysis have not been discussed, or have been discussed in terms of unverifiable assumptions. Here, we attempt to close that gap, focusing on the one-class support vector machine (OC-SVM) method from \cite{Tang2019}. 

\autoref{eqn:diss} may be rewritten for $\mathcal{U}_e\subseteq\mathcal{L}_{2e}$ as
\begin{subequations} \label{eqn:diss_ML}
\begin{align}
\langle \Pi,\Gamma(u) \rangle &\geq 0 \; \forall \; u\in\mathcal{U}_e, \mbox{ where} \\
\Pi &= \bmat{Q & S \\ S^T & R}, \mbox{ and} \\
\Gamma(u) &= \int_0^T \bmat{\mathcal{G}(u(\tau)) \\ u(\tau)}\bmat{\mathcal{G}^T(u(\tau)) & u^T(\tau)}d\tau.
\end{align}
\end{subequations}
Given a set of sample trajectories, $\{u_i,\mathcal{G}(u_i)\}$, with $u_i\in\mathcal{U}_e^r$ and $\mathcal{G}(u_i)\in\mathcal{Y}^m$, $\Gamma(u_i)$ gives a data point in $(m+r)(m+r+1)/2$-dimensional space, and $\Pi$ represents a vector in the same space that defines a half-plane. The hard (not allowing for any categorization error) OC-SVM  solves
\begin{align}
    \min_{\Pi,\rho} \hspace{2mm} \|\Pi\|_F^2/2 - \rho \hspace{2mm} \mathrm{s.t.} \hspace{2mm} \langle \Pi,\Gamma(u_i)\rangle\geq \rho \hspace{2mm} \forall \; u_i\in\mathcal{S}, \label{eqn:OC-SVM}
\end{align}
where $\mathcal{S}$ is the set of sample trajectories, $\rho{\in}\mathbb{R}_+$ measures the distance of the samples from the half-plane, and $\|\Pi\|_F^2/2$ penalizes the boundary complexity \cite{Tang2019}. Equation~\ref{eqn:OC-SVM}
seeks the half-plane closest to the data for which all data appears on only one side. This gives the tightest $(Q,S,R)$ characterization of the data. The question remains whether $\mathcal{S}$ is adequately informative. In probably approximately correct (PAC) learning \cite[\S 3]{ShalevShwartz2014}, the following assumption is used.
\begin{assumption} \label{ass:ML}
    Data, $u_i$, is sampled from $\mathcal{U}_e$ according to a probability distribution, $\mathcal{D}$.
\end{assumption} 
If Assumption~\ref{ass:ML} holds, classifier error is the probability that \autoref{eqn:diss_ML} does not hold with $\Pi$. This is written
    $L_\mathcal{D}(\Pi) \defeq \underset{u_i\sim\mathcal{D}}{\mathbb{P}}\left(\langle \Pi,\Gamma(u_i)\rangle < 0\right).$
On the other hand, the empirical loss is 
    $L_\mathcal{S}(\Pi) \defeq |\{\mathcal{S}\, | \, \langle \Pi,\Gamma(u_i)\rangle < 0\}|/|\mathcal{S}|.$
For noiseless samples, $\{u_i,\mathcal{G}(u_i)\}$, of a $(Q,S,R)$-dissipative system, $\Pi(Q,S,R)$ results in $L_\mathcal{S}(\Pi) = 0$ by definition, though it may be appropriate to allow for nonzero loss for noisy data. More importantly, the generalization error, $|L_\mathcal{D}-L_\mathcal{S}|$, measures how well $\Pi$ generalizes from $\mathcal{S}$ to the rest of $\mathcal{U}_e$. Assuming $L_\mathcal{S} = 0$, a bound with probability $1-\delta$ on the generalization error of OC-SVM is given in \cite{Scholkopf2001} as
\begin{align*}
    \underset{u_i\sim\mathcal{D}}{\mathbb{P}} &\left(\langle \Pi,\Gamma(u_i)\rangle < \rho - \gamma \right) \\ &\leq \frac{2}{K}\left(\log_2\left(\frac{K^2}{2\delta}\right)+\frac{16c^2}{\hat{\gamma}^2}\log_2\left(\frac{\ln(2)}{4c^2}\hat{\gamma}^2K\right) +2\right),
\end{align*}
where $K$ is the number of samples, $c = 103$, $\hat{\gamma} = \gamma/\|\Pi\|_F$, and $\gamma\in\mathbb{R}_+$ is a parameter that relaxes the offset distance $\rho$ to allow for data points closer to the boundary of the half-plane. Since only $\langle \Pi,\Gamma(u_i)\rangle \geq 0$ is required, $\gamma=\rho$ should be chosen to minimize the number of samples necessary. It's noted in \cite{Scholkopf2001} that $c$ may be 50 times smaller in practice, and tighter bounds may have become available since the original work. In any case, this provides a calculable probabilistic guarantee that future samples will be accurately characterized. Importantly, this bound does not scale with the number of basis functions of $\mathcal{U}_{A1234}$ or with the input and output dimensions. This generalization error is only achieved if the data is sampled according to distribution $\mathcal{D}$, the meaning of which is not immediately clear.

\begin{figure}
    \centering
    \begin{tikzpicture}

\def\opacity{.2}
\def\mywidth{1.5}

\begin{axis}[%
at={(0\textwidth,0\textwidth)},
width=0.6\columnwidth,
height=0.4\columnwidth,
scale only axis,
separate axis lines,
every outer y axis line/.append style={black},
every y tick label/.append style={font=\color{black}},
xmin=0,
xmax=1,
ymin=0,
ymax=1.25,
ylabel ={$\|G(u)\|_T$},
xlabel={$\|u\|_T$},
ylabel style={yshift=-.25cm},
legend style={legend cell align=left, align=left, draw=white!15!black, font = \footnotesize},
legend pos = outer north east,
xtick = {0, .25, .5, .75, 1},
ytick = {0, .25, .5, .75, 1, 1.25}
]

\addplot [color=black, line width=\mywidth pt]
table[row sep=crcr]{%
	0 0	\\
	1 .25 \\
};
\addlegendentry{True}

\addplot[name path=HundredLow,pink,domain={0:1}, line width=\mywidth pt] {.2868*x} node[]{};
\addlegendentry{$b = 100$}

\addplot[name path=TenLow,green,domain={0:1}, line width=\mywidth pt, dashed] {.5049*x} node[]{};
\addlegendentry{$b = 10$}

\addplot[name path=FourLow,red,domain={0:1}, line width=\mywidth pt, dashdotted] {.7473*x} node[]{};
\addlegendentry{$b = 4$}

\addplot[name path=TwoLow,blue,domain={0:1}, line width=\mywidth pt, dotted] {.9874*x} node[]{};
\addlegendentry{$b = 2$}

\addplot [color=black, line width=\mywidth pt]
table[row sep=crcr]{%
	0 0	\\
	1 1.25 \\
};

\addplot[name path=HundredHigh,pink,domain={0:1}, line width=\mywidth pt, dotted] {1.2069*x} node[]{};

\addplot[color=pink, opacity=\opacity]fill between[of=HundredLow and HundredHigh, soft clip={domain=0:1}]
;

\addplot[name path=TenHigh,green,domain={0:1}, line width=\mywidth pt, dashed] {1.1997*x} node[]{};

\addplot[color=green, opacity=\opacity]fill between[of=TenLow and TenHigh, soft clip={domain=0:1}]
;

\addplot[name path=FourHigh,red,domain={0:1}, line width=\mywidth pt, dashdotted] {1.1999*x} node[]{};

\addplot[color=red, opacity=\opacity]fill between[of=FourLow and FourHigh, soft clip={domain=0:1}]
;

\addplot[name path=TwoHigh,blue,domain={0:1}, line width=\mywidth pt, dotted] {1.2*x} node[]{};

\addplot[color=blue, opacity=\opacity]fill between[of=TwoLow and TwoHigh, soft clip={domain=0:1}]
;

\end{axis}

\end{tikzpicture}
    \vspace{-.8cm}
    \caption{Bounds on 1000 trajectories of \autoref{eqn:example} uniformly sampled from $\mathcal{U}_{A1234}$ with $T=20$ and different numbers of basis functions, $b$. For each case of $b$, the upper bound is identical.} \label{fig:ML}
\end{figure}
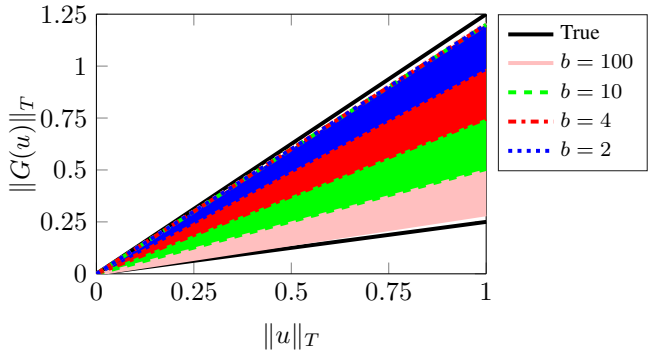

One interpretation is to sample uniformly from $\mathcal{U}_{A1234}$, which supposes that every such signal is equally likely to occur. This is essentially the perspective taken in \cite{Tang2021}, where input trajectories are generated by uniformly sampling the bounded coefficients ($\alpha_i$ in Assumption 4) of $b$ Fourier bases. Since a $\delta$-covering is not needed to achieve confidence in the results, and since the generalization bound does not grow with the data dimension, a large number of basis functions can be used to represent $\mathcal{U}_{A1234}$, making it a better approximation of $\mathcal{L}_{2e}$. This is demonstrated in Figure~\ref{fig:ML}, which depicts the upper and lower bounds of 1000 trajectories of \autoref{eqn:example} randomly sampled from $\mathcal{U}_{A1234}$ with different numbers of basis functions over $T=20$. As with the $\delta$-covering method, a long time horizon results in a good characterization of the upper conic bound (about 4\% error in every case), and more basis functions result in a better characterizations of the lower conic bound (almost 300\% error with $b=2$, and only 14.7\% error with $b=100$). The difference is that adding these basis functions does not increase the sample complexity, so a small generalization error is practically attainable with $b=100$ or even greater. Notably, the number of samples in \autoref{fig:ML} is too small to apply the generalization error bound, yet the data with $b=100$ already covers the entire cone well.


\begin{figure}
    \centering
    \begin{tikzpicture}

\def\opacity{.3}
\def\mywidth{1.5}

\begin{axis}[%
at={(0\textwidth,0\textwidth)},
width=0.6\columnwidth,
height=0.4\columnwidth,
scale only axis,
separate axis lines,
every outer y axis line/.append style={black},
every y tick label/.append style={font=\color{black}},
xmin=0,
xmax=1,
ymin=0,
ymax=1.25,
ylabel ={$\|G(u)\|_T$},
xlabel={$\|u\|_T$},
ylabel style={yshift=-.25cm},
legend style={legend cell align=left, align=left, draw=white!15!black, font = \footnotesize},
legend pos = outer north east,
xtick = {0, .25, .5, .75, 1},
ytick = {0, .25, .5, .75, 1, 1.25}
]

\addplot [color=black, line width=\mywidth pt]
table[row sep=crcr]{%
	0 0	\\
	1 .25 \\
};
\addlegendentry{True}

\addplot[name path=PtOneLow,blue,domain={0:1}, line width=\mywidth pt, dotted] {.2501*x} node[]{};
\addlegendentry{$T = .1$}

\addplot[name path=OneLow,red,domain={0:1}, line width=\mywidth pt, dashdotted] {.2544*x} node[]{};
\addlegendentry{$T = 1$}

\addplot[name path=TenLow,green,domain={0:1}, line width=\mywidth pt, densely dashed] {.4385*x} node[]{};
\addlegendentry{$T = 10$}

\addplot[name path=HundredLow,pink,domain={0:1}, line width=\mywidth pt,loosely dashed] {1.0325*x} node[]{};
\addlegendentry{$T = 100$}

\addplot [color=black, line width=\mywidth pt]
table[row sep=crcr]{%
	0 0	\\
	1 1.25 \\
};

\addplot[name path=PtOneHigh,blue,domain={0:1}, line width=\mywidth pt, dotted] {0.2978*x} node[]{};

\addplot[color=blue, opacity=\opacity]fill between[of=PtOneLow and PtOneHigh, soft clip={domain=0:1}]
;

\addplot[name path=OneHigh,red,domain={0:1}, line width=\mywidth pt, dashdotted] {0.6054*x} node[]{};

\addplot[color=red, opacity=\opacity]fill between[of=OneLow and OneHigh, soft clip={domain=0:1}]
;

\addplot[name path=TenHigh,green,domain={0:1}, line width=\mywidth pt, densely dashed] {1.1802*x} node[]{};

\addplot[color=green, opacity=\opacity]fill between[of=TenLow and TenHigh, soft clip={domain=0:1}]
;

\addplot[name path=HundredHigh,pink,domain={0:1}, line width=\mywidth pt, loosely dashed] {1.2467*x} node[]{};

\addplot[color=pink, opacity=\opacity]fill between[of=HundredLow and HundredHigh, soft clip={domain=0:1}]
;

\end{axis}

\end{tikzpicture}
    \vspace{-.8cm}
    \caption{1000 trajectories of \autoref{eqn:example} from Weiner processes with different lengths, $T$. The Weiner process is implemented as a discrete random walk with time step $0.01$ and step size from a standard normal distribution.} \label{fig:dithering}
\end{figure}
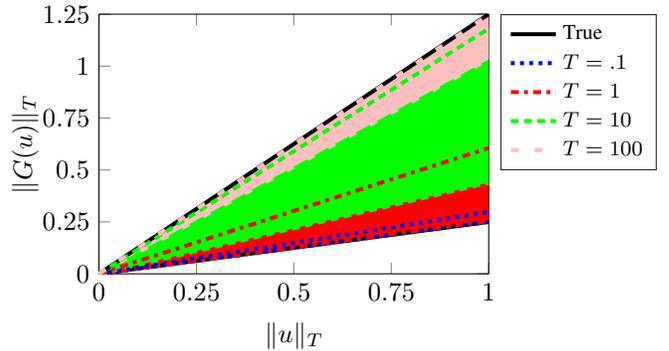

\begin{figure}
    \centering
    \input{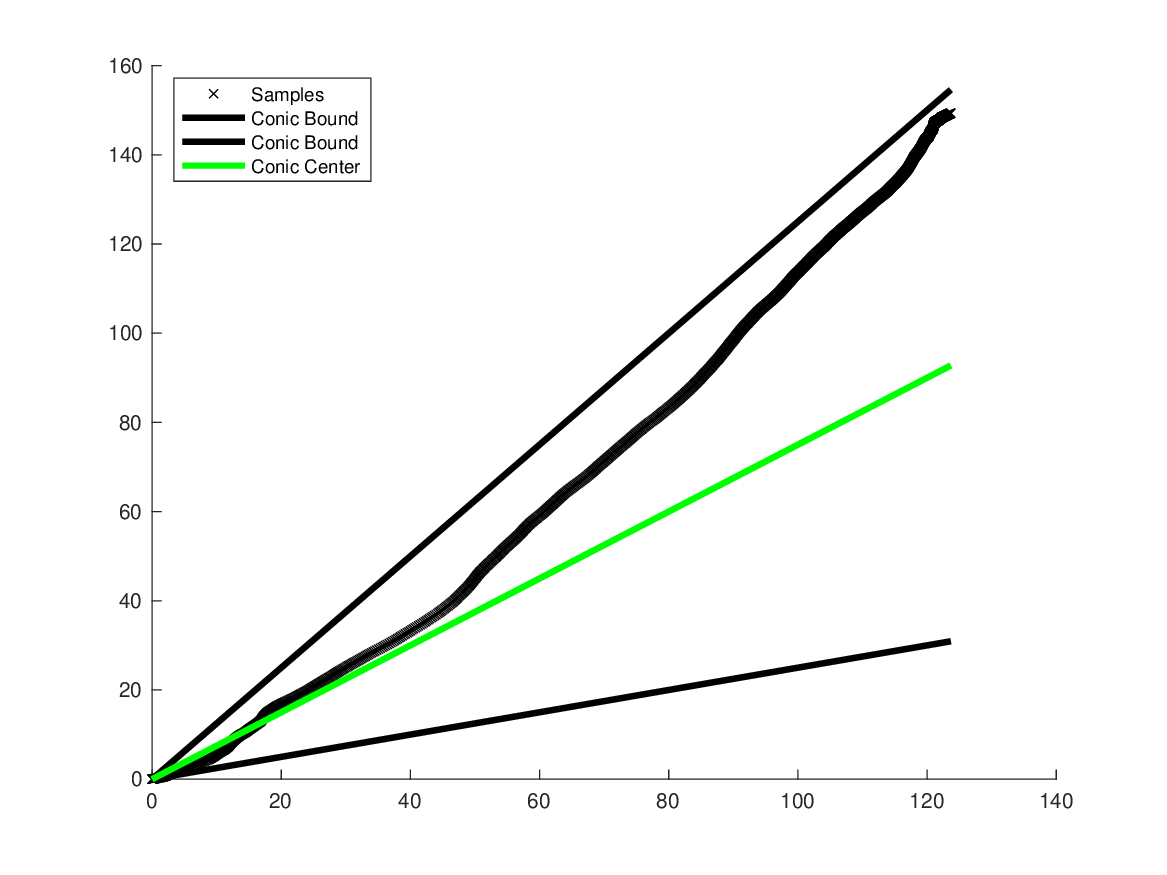}
    \vspace{-.5cm}
    \caption{Data generated from a single Weiner process applied to Equations~\ref{eqn:example} and \ref{eqn:nonlinear} with $T {=} 0.2$ to $4$ (left), and $T{=}0.2$ to $50$ (right) incremented by $0.2$.} \label{fig:dithering2}
\end{figure}

A second interpretation is to randomly generate input signals from a Weiner process. This is the perspective implicitly taken in \cite{Tang2019,Tang2019journal}. A Weiner process is the limit of a random walk, so at each time step, the Weiner process samples randomly from all possible next time steps. This approach does not require Assumptions 1, 2, or 4. However, the Weiner process has a tendency to move away from its origin over time, which amplifies the contribution of low-frequency information. As shown in Figure~\ref{fig:dithering}, the system's response converges to its zero-frequency response as $T\rightarrow\infty$, failing to characterize the lower conic bound for \autoref{eqn:example}. This trend is in conflict with the justification of Assumption 3, which suggests that $T$ should be chosen as large as possible to approximate ultimate virtual dissipativity. One solution to this is to use Weiner processes with uniformly sampled time lengths, $T\in[T_\mathrm{min},T_\mathrm{max}]$. Or, more efficiently, calculate the dissipativity in response to a single or a few Weiner processes for all $T=\Delta t,2\Delta t,\dots,T_\mathrm{max}$, with some increment $\Delta t$. Either way, Assumption 3 can be removed, and dissipativity can be estimated directly instead of inferred from ultimate virtual dissipativity. This is depicted in Figure~\ref{fig:dithering2}, which shows short Weiner process inputs invoke high-frequency response, while long Weiner process inputs invoke low-frequency response. For large enough $T_\mathrm{max}$ and small enough $T_\mathrm{min}$, the cone is well characterized by the data.

Choosing $u_i$ to be Weiner processes with different lengths appears to work exceptionally well for the simple examples studied here. However, this result should be interpreted cautiously when extending to general nonlinear systems. A Weiner process results in an input signal that is persistently exciting, so this is effectively a circuitous application of Willems' Fundamental Lemma when applied to linear systems \cite{Willems2005}. Nonetheless, the interpretation of a Weiner process sampling from the underlying distribution, $\mathcal{D}$, of signals in $\mathcal{L}_{2e}$ is an appealing extension to nonlinear systems. 

\section{Conclusions}
To use dissipativity for stability analysis, it must hold on a realistic set of input signals, which is usually $\mathcal{L}_{2e}$ in reality. However, data can only practically be collected from a finite subset of $\mathcal{L}_{2e}$, so ensuring that the estimated dissipativity properties generalize from the dataset to $\mathcal{L}_{2e}$ is a major challenge. This work demonstrated that $\delta$-coverings require a sample size that becomes intractable at a super-exponential rate as the sample region approaches $\mathcal{L}_{2e}$, whereas machine learning  can achieve a probabilistic generalization error that does not increase as the sample region approaches $\mathcal{L}_{2e}$.

\section{Appendix}
In \cite{Montenbruck2016}, it is shown that a system, $\mathcal{G}:\mathcal{U}\rightarrow\mathcal{Y}$, is virtually $(-I,0,\gamma I)$-dissipative for some $\gamma$ satisfying
\begin{align*}
\gamma \leq \max_{u'\in\mathcal{U}' : \|u'\|\geq \delta} \frac{L\delta + \|\mathcal{G}(u')\|}{\|u'\|-\delta},
\end{align*}
where $L$ is the Lipschitz constant of $\mathcal{G}$, and $\delta$ is the covering radius of $\mathcal{U}'$ on $\mathcal{U}_{A1234}$. A similar expression was derived for virtual $(lI,I,0)$-dissipativity. Both of these calculations incur $\mathcal{O}(1)$ space complexity. A method for estimating conic sectors (i.e. $(-I,c,r^2-c^2)$-dissipativity with the minimal $r>0$) is also proposed, but it incurs $\mathcal{O}(K)$ space complexity due to the least squares estimation of $c$, which is also unnecessarily susceptible to sampling bias. Here, we derive expressions with $\mathcal{O}(1)$ space complexity for the tightest upper, $b$, and lower, $a$, conic bounds satisfying $(-I,\frac{a+b}{2}I,-abI)$-dissipativity. For conic bounds $a$ and $b$, Equation~\ref{eqn:diss} may be rewritten
\begin{align}
    \bmat{\|\mathcal{G}(u)\|_T^2 & \langle \mathcal{G}(u),u\rangle_T & \|u\|_T^2}R_{ab} \geq 0, \label{eqn:diss_mat}
\end{align}
where $R_{ab} = [-1, \; a{+}b, \; -ab]^T$. Letting $L$ be the Lipschitz constant of $\mathcal{G}$ and $\delta$ be the covering radius of $\mathcal{U}'_{A1234}$ on $\mathcal{U}_{A1234}$, it can be shown (with similar arguments to \cite{Romer2017}), if
\begin{align}
    M^T(u_i)R_{ab} \geq 0 \label{eqn:diss_mat_robust}
\end{align}
is satisfied for all $u_i\in\mathcal{U}'_{A1234}$, then Equation~\ref{eqn:diss_mat} is satisfied for all $u\in\mathcal{U}_{A1234}$,
where $M(u_i) \defeq [(\|\mathcal{G}(u_i)\|_T \pm \Lambda_2)^2 , \; \langle \mathcal{G}(u_i),u_i\rangle_T \pm \Lambda_3 , \; (\|u_i\|_T \pm \Lambda_1)^2]$,
and $\Lambda_1 \defeq \delta$, $\Lambda_2 \defeq L\delta$, and $\Lambda_3 \defeq L\delta\|u_i\|_T + \delta\|\mathcal{G}(u_i)\|_T + L\delta^2$ come from bounding $| \|u\|_T - \|u_i\|_T|$, $| \|\mathcal{G}(u)\|_T - \|\mathcal{G}(u_i)\|_T|$, and $|\langle \mathcal{G}(u),u\rangle_T - \langle \mathcal{G}(u_i),u_i\rangle_T|$ with the triangle and Cauchy-Schwartz inequalities, recalling that $\forall u{\in}\mathcal{U}_{A1234}$, $\exists u_i{\in}\mathcal{U}'$ where $\|\mathcal{G}(u)-\mathcal{G}(u_i)\|_T{\leq} L\|u-u_i\|_T$ and $\|u-u_i\|_T{\leq} \delta$.

There are two ways to find the tightest upper and lower bounds \cite{BridgemanPassivity}. One option is to maximize the lower bound, denoted $a_\mathrm{R}$, then minimize the corresponding upper bound, denoted $b_\mathrm{I}$. The other is to minimize the upper bound, denoted $b_\mathrm{L}$, then maximize the corresponding lower bound, denoted $a_\mathrm{I}$. To find $a_R$, divide Equation~\ref{eqn:diss_mat_robust} by $b$ and let $b\rightarrow \infty$. This results in 
$
    \langle \mathcal{G}(u_i),u_i \rangle_T \pm \Lambda_3 \geq a_R (\|u_i\|_T \pm \Lambda_1)^2.$
Therefore, the largest possible value of $a_R$ is
\begin{align}
    a_R = \min_{\{i \, | \, \|u_i\|_T -\Lambda_1> 0\}} \frac{\langle \mathcal{G}(u_i),u_i\rangle_T - \Lambda_3}{(\|u_i\|_T \pm \Lambda_1)^2}. \label{eqn:aR}
\end{align}
Following similar reasoning to derive $b_\mathrm{I}$, let $q=-1$, $s = (a_\mathrm{R}+b_\mathrm{I})/2$, and $r = -a_\mathrm{R}b_\mathrm{I}$. Rearranging yields $N_2 \leq b_\mathrm{I}D_2$, where
    $N_2\defeq \|\mathcal{G}(u_i)\|_T^2 + \Lambda_2 - a_\mathrm{R}(\langle \mathcal{G}(u_i),u_i\rangle_T \pm \Lambda_3)$, and
    $D_2 \defeq \langle \mathcal{G}(u_i),u_i\rangle_T \pm \Lambda_3 - a_\mathrm{R}(\|u_i\|_T\pm\Lambda_1')^2$.
Here, $D_2$ is positive or negative depending on $a_\mathrm{R}$ and $\langle \mathcal{G}(u_i),u_i\rangle_T$. Since the minimum consistent value of $b_\mathrm{I}$ is desired, rearranging the equation should bound it below. Dividing by $D_2<0$ yields an upper bound, which is extraneous, and dividing by zero is singular. Therefore, only samples satisfying $D_2>0$ must be considered. For these values, $b_\mathrm{I}\geq N_2/D_2$. Therefore,
$b_\mathrm{I} = \max_{\{i|D_2>0\}} N_2/D_2$. Similar arguments result in
\begin{align}
    b_\mathrm{L} &= \max_{\{i \, | \, \|u_i\|_T - \Lambda_1 > 0\}} \frac{\langle G(u_i),u_i\rangle_T - \Lambda_3}{(\|u_i\|_T \pm \Lambda_1)^2}, \label{eqn:bL} 
\end{align}
and $a_\mathrm{I} {=} \max_{\{i|D_1>0\}} N_1/D_1$, where $N_1 {=} b_\mathrm{L}(\langle \mathcal{G}(u_i),u_i\rangle_T\pm \Lambda_3) - \|\mathcal{G}(u_i)\|_T^2 + \Lambda_2$ and $D_1 = b_\mathrm{L}(\|u_i\|_T \pm \Lambda_1')^2 - (\langle \mathcal{G}(u_i),u_i\rangle_T \pm \Lambda_3)$. These bounds always obey the relation $a_\mathrm{I}\leq a_\mathrm{R}\leq b_\mathrm{L}\leq b_\mathrm{I}$, and $\mathcal{G}$ is in $\mathrm{cone}(a_\mathrm{I},b_\mathrm{L})$, $\mathrm{cone}(a_\mathrm{R},b_\mathrm{I})$, $\mathrm{cone}(a_\mathrm{R},\infty)$, and $\mathrm{cone}(-\infty,b_\mathrm{L})$.


\addtolength{\textheight}{-12cm} 
\bibliographystyle{IEEEtran}
\bibliography{IEEEabrv,Bibliography}

\end{document}